\documentclass[12pt]{article}

\usepackage{scicite}
\usepackage{times}
\usepackage{graphicx}
\usepackage{caption}
\usepackage[nomarkers]{endfloat}

\newenvironment{sciabstract}{%
\begin{quote} \bf}
{\end{quote}}

\title{A New Information Theory of Certainty for Machine Learning}

\author
{Arthur Jun Zhang, $^{1,2}$\\
\\
\normalsize{${^1}$ Jayoo Technology LLC,}\\
\normalsize{ 21 Fainwood Circle, Cambridge, MA 02139, USA}\\
\normalsize{$^{2}$ Department of Mathematics, Brandeis University,}\\
\normalsize{415 South Street, Waltham, MA 02453, USA}\\
\\
\normalsize{E-mail:  arthurjunzhang@gmail.com.}
}

\date{}

\begin{document} 


\baselineskip24pt


\maketitle


\begin{sciabstract}
  Claude Shannon coined entropy to quantify the uncertainty of a random distribution for communication coding theory in 1948\cite{shannon}. Ever since then, entropy has been widely used and equivalently treated as {\it information\/} by many people, and Shannon information theory has achieved huge success in all science and engineering fields. However, we observe that the uncertainty nature of entropy also limits its direct usage in mathematical modeling such as  prediction tasks. Entropy has been mainly used as a measure of information gain or loss and it hardly has been directly employed for modeling and prediction purposes in the literature. We notice that a quantity measuring the certainty of a random distribution is the desired {\it directly usable\/} information for such modeling purposes. Therefore we propose a new information concept {\it troenpy\/}, as the canonical dual of entropy, to quantify the certainty of the underlying distribution. We establish the necessary concepts and properties for this new Information Theory of Certainty (ITC), analogue of the classical Shannon information theory.   We demonstrate two important applications of troenpy for machine learning. The first is for the classical supervised document classification task, we develop a troenpy based weighting scheme to leverage the document class label distribution and show that the weighting scheme can be easily and very effectively used for classification tasks. The second is for the popular self-supervised language modeling task, where we introduce a self-troenpy weighting scheme for sequential data and show that it can be directly and easily included in modern recurrent neural network based language models and achieve dramatic perplexity reduction. Besides machine learning ITC also has the potential application on quantum information theory. We generalize the idea and define {\it quantum troenpy\/} as the dual of the Von Neumann entropy to quantify the certainty of quantum systems. In conclusion we developed a new information theory quantifying certainty in random systems and such information can be easily, directly and effectively used in modern machine learning and neural network models. With the theory supported cheap and effective way of extracting and representing useful information from data, ITC offers not only promising ways to leverage useful information to improve current machine learning model performances but also possibilities of designing new machine learning and neural network models from the view of useful information processing.
  
\end{sciabstract}


\section*{Introduction}

Claude Shannon first invented entropy to measure the uncertainty of transported message errors in communication channels\cite{shannon}. It is the central concept in the famous Shannon coding theorems, which founded the communication industry with a rigorous mathematical theory. 
The success of Shannon information theory happened in not only communication industry but also all science and engineering fields. It has been widely used as a mathematical tool characterizing uncertainty in random systems. Ever since then, entropy has been widely treated equivalently as {\it information\/} itself in the science community.  Researchers often compute the change of entropy as the information gain or loss for uncertainty comparison and decision makings purposes. For example, the decision tree algorithms in machine learning usually employ entropy as the default metric for information gain on tree node splitting decisions.  The maximum entropy principle \cite{Jaynes} is often used as a guiding principle for model optimization when the training data is regarded as the testable prior information. The generalized concept cross entropy is widely used as the standard objective loss function for categorical random variables in machine learning nowadays.

However, entropy hardly has been directly included into mathematical expressions for model computations. We hypothesize this is due to the intrinsic nature of uncertainty that entropy measures. In a sense it is a type of {\it negative information\/} measuring the disorder in a random distribution. For comparison of uncertainty levels it works perfectly fine. But we cannot directly make use of such information measurement for mathematical modeling  of prediction tasks, or at least entropy is not the optimal candidate for such purposes. The reciprocal of entropy may work for some cases, but entropy may not the ideal candidate.

Also in artificial intelligence community, leading research groups have recently invested heavily on building large scale neural network models such as the popular GPT language models\cite{gpt2018,gpt2019,gpt2020,gpt42023}.  These giant models are trained with massive amount of data using modern optimization algorithms such as Adam and variations of stochastic gradient descent. All the useful information and knowledge in the data are {\it only\/} learned through the gradient descent optimization and  are stored in the model parameters together with other unwanted noise information. This explains nowadays the state-of-the-art neural network models for various tasks have become larger and larger in order to perform better and be more knowledgeable. It would be ideal if one could directly extract some useful global information from the raw data and can later transfer and integrate these information into other downstream models. This way the models can benefit from the precomputed information and achieve substantial performance improvement.

With the idea in mind, our recent investigation starts with the well-known weighting scheme Inverse Document Frequency (IDF)\cite{idf1972}. For a given word $w$, it is simply given as IDF($w$)=$1+log(n/(1+d))$, where $n$ is the number of documents in a corpus and $d$ is the number of documents containing the word $w$.  It is simple but effective. We were very motivated and started exploring if other weighting schemes exists. The connection to information theory pointed out by Aizawa\cite{aizawa2003} finally leaded us to Shannon entropy. The IDF can be interpreted as the {\it self-information\/} of sampling a term in a document in the sense of information theory and TF-IDF can be interpreted as the mutual information between random sampling a document and random sampling a term in a document. For a discrete random variable $X$ with $P(X=v_i)=p_i$, where $i\in \{1,2,\dots,K\}$,  Shannon uses the {\it self-information\/} quantity $log(1/p_i)$ measures  the {\it surpriseness \/} and the entropy is  therefore interpreted as the expected {\it rareness\/}.

The critical thinking is that since the rareness approach works well, observed by the huge success of Shannon information theory, can the {\it commonness\/} approach, regarded as the contrary of rareness, work?  
This simple idea leads us to the discovery of a new notion of information \cite{troenpy, usptoTroenpy} and we name it {\bf Troenpy} as a natural dual to entropy. Note the entropy measures the rareness and the self-information is also called {\it negative information}, the entropy is often interpreted as the measurement of uncertainty. Correspondingly, our analogue of self-information is called {\it positive information\/} and it is defined as $log(1/(1-p_i))$. Our novel information quantity troenp measures the certainty of a distribution, or interpreted as the commonness in a distribution. They two naturally are twins and form a family. They complement each other and each has its own expertise on the applications. Entropy measures uncertainty  and excels at error analysis while troenpy quantifies certainty and excel at mathematical modeling and prediction tasks. For some tasks such as comparison and decision making, they are equally good and give identical results when used as a metric of information gain or loss.

Certainty of a random variable measures the reliability of the underlying distribution. After all, mathematical modeling relies on the certainty of model inputs to say something about the outputs with confidence. Therefore we believe the information of certainty is the ideal and desired  type of information to use for mathematical modeling. This type of information can be first precomputed from datasets and later be directly included in mathematical models for relevant downstream tasks.  In particularly, the relevant global information of the data would effectively guide the landscape of the model parameters during model training and  substantially improve the overall performance. The only similar feature that  has  been used before is the IDF.

In the following sections we will first define troenpy and establish the basic concepts and properties for ITC. Then we will introduce a troenpy based  weighting method for the classical supervised documents classification task, which  leverages of the documents class distribution and can dramatically lower the classification errors of downstream models. For the popular self-learning on sequential data without label information, we introduce another weighting scheme, namely self-troenpy. We illustrate that self-troenpy can be easily integrated into recurrent neural network based language models and very effectively achieve perplexity reduction on two benchmark datasets for language modeling.  Finally for the potential application on quantum information theory, we define the concept of quantum troenpy for quantum information theory of certainty.

\section*{Essentials of Information Theory for Certainty}
In this section we give the basic concepts necessary to establish the Information Theory of Certainty (ITC). Let's fix the nations first. We let $X$ denote a discrete random variable with values in a finite space $\chi=\{v_1,v_2,\dots,v_K\}$ and the probability mass function $p(X=v_i)=p_i$ for $i \in \{1,2,\dots,K\}$. Recall the self-information $\textrm{log}(1/p_i)$ measures the surpriseness  for $x$ taking the value $v_i$ and because of the uncertainty nature, we call it \emph{negative information}.  The expression $-p_i\textrm{log}p_i$ is the contributed uncertainty or surpriseness for $x$ taking the value $v_i$. Naturally the expression   $\textrm{log}(1/(1-p_i))$ measures the commonness or non-surpriseness level for $x$ taking the value $v_i$, and it is called the positive information. The term $-p_i\textrm{log}(1-p_i)$ measures the  certainty contribution for $x$ taking $v_i$. We define {\it {\bf troenpy}\/} to be the expectation across all possible values, denoted as  $$\textrm{T}(X):=-\sum^{K}_{i=1}p_i\textrm{log}(1-p_i)$$ Troenpy is the integral of positive information and thus it measures the certainty of $X$ across the whole distribution. For mathematical modeling, the higher the troenpy is, the more reliability the variable has, thus more prediction capacity it has. Note also the function $t(x)=-x\textrm{log}(1-x)$ is a convex function for $x \in (0,1)$.

Note that the sum  $\sum_{i=1}^K(1-p_i)=\textrm{K}-(p_1+\dots,p_K)=\textrm{K}-1$. If we let $q_i=(1-p_i)/(\textrm{K}-1)$, then $q=(q_1,\dots,q_K)$ is a probability distribution. According to the Gibbs inequality, the cross entropy $-\sum_{i=1}^Kp_ilogq_i$ achieves minimum value when $p_i=q_i$, which immediately gives $p_i=1/\textrm{K}$. This shows that the troenpy achieves minimum value when the distribution is evenly distributed across the value set, while entropy achieves the maximum value.  It is also obvious that the troenpy can be treated as the above cross entropy plus a constant $log(\textrm{K}-1)$.

 Let $p(x,y)$ denote the joint distribution of the random variables $X$ and $Y$, and lowercase letters denote the random variable values. We define the {\bf conditional troenpy } of $X$ given $Y$, denoted as  $\textrm{T}(X|Y)$,  to be the following $\textrm{T}(X|Y)=\sum_yp(y)\textrm{T}(X|Y=y)$. It can further be reduced to $\textrm{T}(X|Y)=-\sum_{y,x}[p(y)p(x|y)\textrm{log}(1-p(x|y))]=-\sum_{x,y}p(x,y)\textrm{log}(1-p(x|y)).$

We define the {\bf Pure Positive Information} of $X$ from knowing $Y$, denoted as $\textrm{PPI}(X;Y)$,  to be the troenpy gain below:$$\textrm{PPI}(X;Y)=\textrm{T}(X|Y)-\textrm{T}(X)=\sum_{x,y}p(x,y)\textrm{log}\frac{1-p(x)}{1-p(x|y)}$$

\noindent Note in general $\textrm{PPI}(X;Y)\neq \textrm{PPI}(Y;X)$. This is very different from the mutual information $\textrm{MI}(X;Y)$ of two random variables $X$ and $Y$ in the literature, where $\textrm{MI}(X;Y)=\textrm{MI}(Y;X)$. In order  for them to be equal, this requires $(1-p(x))/(1-p(x|y))=(1-p(y))/(1-p(y|x))$, which is equivalent to $p(x)-p(x|y)=p(y)-p(y|x)$. However, this last equation does not hold in general. Analogous to the Shannon information theory, other properties and related maximum principles can be correspondingly established.

\section*{Weighting Scheme Gifts for Machine Learning and Neural Networks}
Using the established ITC, in this section we show how the troenpy can be used to extract useful information from raw dataset, which can be later directly used as weighting schemes for text document classification task and language modeling. We will also compute three versions of  entropy based weightings for comparison and validate the claim that troenpy is a lot more effective than entropy as a weighting scheme. We demonstrate two common scenarios to extract troenpy information. The first is when the class labels of documents is available, which is typical for classification tasks. The second scenarios is when the class label information is not available, where the task is often reformulated as a self-learning task nowadays. Before we go to the details, let us first answer and explain the following question.

\noindent {\bf Why we are interested in weighting schemes ?}  Our original idea comes from the well-known and widely used Inverse Document Frequency(IDF) weighting scheme \cite{idf1972} for information retrieval. In the retrieval tasks one simply multiplies the Term Frequency(TF) with IDF for a feature. It is simple but very effective. The whole ITC is developed for achieving this goal. It is the simplest and economical way we can hope for using information efficiently.

\subsection*{PCF Weighting Scheme for Document Class Label Information}
In this section we consider the scenario when the documents' label information is available,  and we show how to make use of them using the established ITC. Note before this work, in the literature of  machine learning people did not know how to make systematic use of such information and include into the models for performance improvement. First let's fix some notations. We use $\textrm{DC}_{*}$ denote the whole documents collection of the corpus.  For a specific word $w$, we use  $\textrm{DC}_1$ denote all the documents collection with the word $w$ presence in the corpus. Similarly, we use  $\textrm{DC}_0$ denote all the documents collection with the word $w$ absence in the corpus. We denote  the total number of different document class labels as $\textrm{K}$. Then for $\textrm{DC}_*$, we count the number of documents for each class $c_i$, denoted as $\textrm{ct}_i$, where $i \in \{1,2,\dots,\textrm{K}\}$.  For the count statistic $\textrm{CT}=[\textrm{ct}_1,\dots,\textrm{ct}_K]$, we normalize to a probability distribution and denote the troenpy of $\textrm{CT}$ as $\textrm{T}(\textrm{DC}_*)=\textrm{Troenpy}(\textrm{CT})$. The troenpy $\textrm{T(DC}_*)$ measures the overall certainty level of class label distribution in the whole collection $\textrm{DC}_*$ across all words.

Similarly, for a specific word $w$ we compute the troenpy for the document collection $\textrm{DC}_1$ with the word $w$ present.  Then we count the class label statistics for $\textrm{DC}_1$, denoted as $\textrm{CT}_1$, and normalize it to a probability distribution. The corresponding troenpy of normalized $\textrm{CT}_1$ is denoted as $\textrm{T}(\textrm{DC}_1)$, which measures the certainty level of class label distribution for the collection $\textrm{DC}_1$. The {\bf Positive Class Frequency(PCF)} is simply defined to be the difference of the two troenpies
$$\textrm{PCF}(w):=\textrm{T(DC}_1)-\textrm{T(DC}_*)$$

\noindent Without abuse of notation, it is just the $\textrm{PPI}(\textrm{DC};w)=\textrm{T(DC}|w)-\textrm{T(DC}_*)=\textrm{T(DC}_1)-\textrm{T(DC}_*)$. The PCF($w$) measures the certainty level of the class label distribution when the word $w$ is present, and we simply include it as the weighting scheme for the term frequency of $w$.

For the documents classification task, we evaluate the PCF weighting in a simple KNN setting. We denote the PCF weighting multiplied with Inverse Document Frequency(IDF) as PI. From the result reported in Figure 1, we observe that the PCF weighting is quite effective and the TF-PI very significantly outperforms the classical TF-IDF with an average of $22.9\%$ error reduction and the popular optimal transportation based word moving distance  method HOFTT with an average of $26.5\%$ error reduction across all seven datasets.

\begin{figure}
  \centering
   \includegraphics[width=0.99\textwidth]{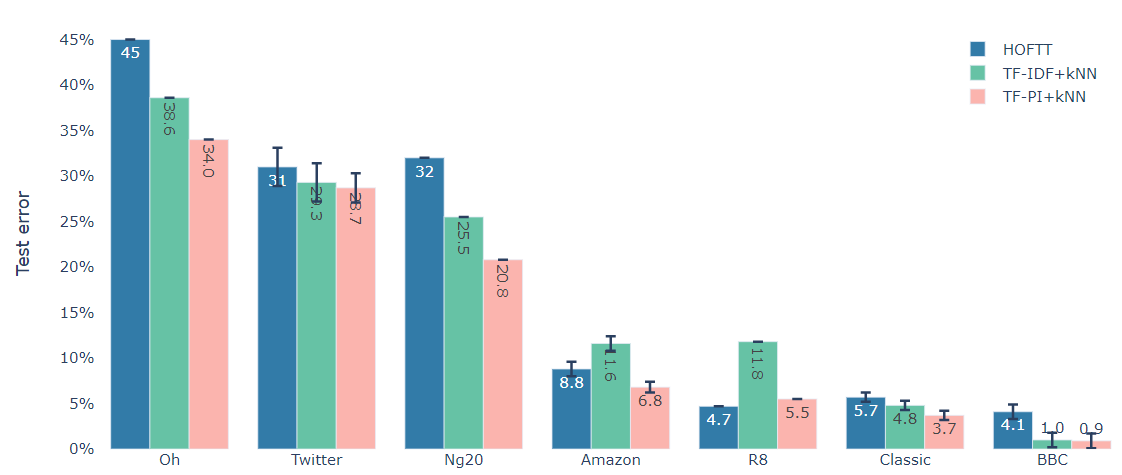}
   \caption{PCF Weighting Scheme based model TF-PI compared with two baseline models the classical TF-IDF and Optimal transportation based Word Moving Distance model HOFTT. The TF-PI model significantly outperforms the two baselines with substantial error reduction.  }
\end{figure}

On the other hand, the entropy is not effective when employed as a weighting scheme. Here we did several experiments to illustrate this fact.
We tried the following experiments to explore the possible combinatorial ways of employing entropy as weightings. For a specific word $w$, we define the {\bf Negative Class Frequency} $\textrm{NCF}_{10}:=\textrm{H(DC}_1)-\textrm{H(DC}_0)$, the entropy difference between document collection with $w$ presence and document collection without $w$. Note here we follow the literature and use $H$ denote the entropy of a distribution. Similarly, $\textrm{NCF}_{1*}:=\textrm{H(DC}_1)-\textrm{H(DC}_*)$ denotes the entropy difference between document collection with  $w$ presence and the whole document collection $\textrm{DC}_*$. And  $\textrm{NCF}_{*0}:=\textrm{H(DC}_*)-\textrm{H(DC}_0)$ denotes the entropy difference between the whole document collection $\textrm{DC}_*$ and the document collection without the word $w$.

In the experiments we respectively used the three NCF weightings instead of the PCF weighting and multiplied with the TF-IDF terms under the same kNN experiment setting as above. Here we used the three datasets with fixed training and test data splits for easy comparison. The result in Figure 2 shows that: first compared with the baseline, the PCF weighting dramatically lowers the error percentages across all datasets , and secondly all the three entropy based weighting schemes are not effective on lowering the error rate across the datasets. This suggests that troenpy based PCF is a better choice as weighting scheme for document representation while entropy based NCF is just not effective for the role.

\begin{figure}
  \centering
   \includegraphics[width=0.99\textwidth]{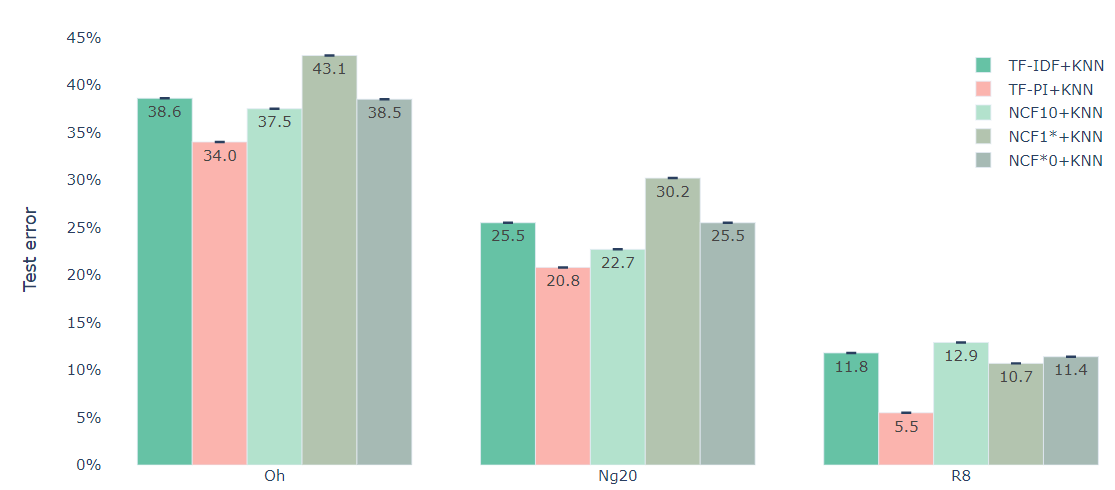}
   \caption{Three Entropy based  Weighting Schemes compared with Troenpy based Weighting Scheme and the Baseline TF-IDF. The troenpy weighting scheme effectively reduces the errors across the three datasets while the entropy based weighting schemes are not effective overall. }
   
\end{figure}

\subsection*{Self-Troenpy Weighting Scheme of Sequential Data for Self-Learning and Language Modeling}

In this section we first introduce Self-Troenpy for sequential data when the class label information we discussed above is not available. And then we show how to compute them. For self-supervised learning, we demonstrate how to integrate these weights into neural network based language models for performance improvement.

{\bf Self-Learning of Sequential Data}
Acquiring document labels usually needs some cost and they are often not available. So let us ignore the document labels for a moment and consider the documents of sequential data only, which is often regarded as  an unsupervised problem. Following the fashion of self-learning in machine learning nowadays, we can reformulate the problem as many simple self prediction tasks. For each word $w_i$ in a sentence sequence $s=[w_0,\dots, w_{i-1}, w_i,w_{i+1},\dots, w_n]$, we can generate a training sample which uses all the words within a window radius of a preselected size $c$ to predict $w_i$. Here suitably large $c$ gives better coverage, for example in the experiments we set  $c=10$. However, too large $c$ will lead to very similar words distribution for each word. This would result in  that many words have roughly the similar weights and thus become less effective. This formulation is the same as that in the famous Continuous Bag Of Words(CBOW)\cite{word2vec}. Now we have  reduced our problem to the above solved supervised problem with label information and one can compute the troenpy of the label distributions the same way as above. The only difference is that our document label space is identical to the feature space of word tokens. 

{\bf Self-Troenpy Definition and Computing}
Now for the computation of troenpy, we do not really need to generate such training samples. This special case is simpler. For each word $w$ in a sentence, it serves as a feature for all the words within a distance $c$ to itself. At the same time it can be predicted by other words within a distance c from it. Now for each sentence $s$, we can scan from the beginning of the sentence to the end and count the statistics for each word in $s$. After scanning and counting the whole corpus of documents, we have all the word counts distribution for each word.
That is, $\textrm{CT}_w=[\textrm{ct}_1,\dots,\textrm{ct}_M]$, where each $\textrm{ct}_i$ is the counts of times that word $w_i$ show up within distance $c$ to the current word $w$ in the corpus, and $M$ is the vocabulary size of the corpus. We normalize $\textrm{CT}_w$ as a distribution and then the troenpy of $\textrm{CT}_w$, namely the Self-Troenpy of $w$, is the positive certainty information about $w$. It says how much certainty or influence the word $w$ has on its neighboring words of size $c$.  Without abuse of notation, we also use  $\textrm{CT}_*=[\textrm{CT}_1,\dots,\textrm{CT}_M]$ denote the total counts of each word in the corpus. After normalizing $\textrm{CT}_*$ to an overall frequency distribution, the troenpy of $\textrm{CT}_*$ measures the certainty level of the whole corpus and it is a constant for every word. In the experiments below we simply use the troenpy of $\textrm{CT}_w$ as the Self-Troenpy of the word $w$, without subtracting the troenpy of $\textrm{CT}_*$.

{\bf Self-Troenpy Weighting Scheme for  Neural Network based Language Models}
Next we show that the Self-Troenpy weighting discussed above can be integrated into language models  and effectively lower the perplexity of language models. We illustrate this by building a simple language model using the standard recurrent neural network LSTM\cite{lstm97}. We use a 2 layers LSTM with 100 hidden units and ReLu as the activation function, a dense layer of also 100 units and a dropout layer with dropout rate 0.2. The output layer is a dense layer of the vocabulary size using the standard softmax as activation function. For the word embedding layers, we set the embedding dimension to be 64. In most neural network frameworks such as Tensorflow or Pytorch embedding vectors are by default  initialized with uniform random numbers from a small interval such as $[-0.1, 0.1]$.    We first normalize each word's embedding vector to norm unit 1 and then multiply the vector with its corresponding precomputed Self-Troenpy weight.

The rational for such weight initialization operation is that the Self-Troenpy measures the influence certainty level of words distribution in the neighborhood of the current word and Self-Troenpy is sort of the prediction capacity norm extracted from the dataset. Higher Self-Troenpy suggests more reliable signals for predicting the next words.  Normalizing embedding vectors to unit vectors and  multiplication with Self-Troenpy is equivalent to asking the underlying optimization algorithm to find the optimal parameters with such initialization constraint.  This is because in the end the learned word embedding vectors should achieve such  influence certainty levels on their neighboring words close to the certainty level computed from the training dataset. This is also well-known in the neural network research community that the initialization of a network parameters is critical for the training phase of networks and has profound impact on the network's performance on downstream tasks\cite{critical18}.
Next our following experiments would validate this argument.

We use the two popular datasets Penn Treebank (PTB) and WikiText-2 (WT2) for evaluation of our new language model. Here we use the perplexity of a language model as the metric \cite{Jurafsky2009}. Perplexity is an intrinsic measure of language models and it can be computed as the exponential of the cross entropy on an evaluation dataset. Lower perplexity suggests better prediction power of the underlying model. We use the training data within each dataset to train the baseline LSTM and Troenpy weighted LSTM models with fixed 30 epochs, and then evaluate on the validation and test datasets respectively. For comparison and reference we also include  an advanced algorithm Mogrifier LSTM \cite{Mogrifier} in the report, a recent state-of-art algorithm. In the Figure 3, we observe that Tro-LSTM very effectively reduces the perplexity of the baseline  LSTM models.  The perplexity reduction is about 5 points  for PTB  and 12 points  for WTB on average over the validation and test datasets. The Tro-LSTM outperforms the Mogrifier LSTM on PTB and  it is also comparable to Mogrifier LSTM on WT2. Note also that one can apply dynamic evaluation trick \cite{dynamicEval} to further lower the perplexity. Note our Tro-LSTM has only 1.7M(million) trainable parameters for PTB and 5.3M parameters for WT2, while Mogrifier LSTM has 24M for PTB and 35M for WT2. 

Interestingly, the troenpy of the corpus words frequency $\textrm{CT}_*$ for both PTB and WT2 are both roughly 0.011, while the entropies for PTB and WT2 respectively equals 6.62 and 7.26. The top 10 words with highest troenpy for both datasets are given in Table 1, where we observe that the WT2 top ranked words have more certainty on neighboring words than the PTB top ranked words. This is consistent with our understanding that the WT2 top words are more specialized in the usage while the PTB top words are relatively common words in news articles.

\begin{figure}
  \centering
   \includegraphics[width=0.99\textwidth]{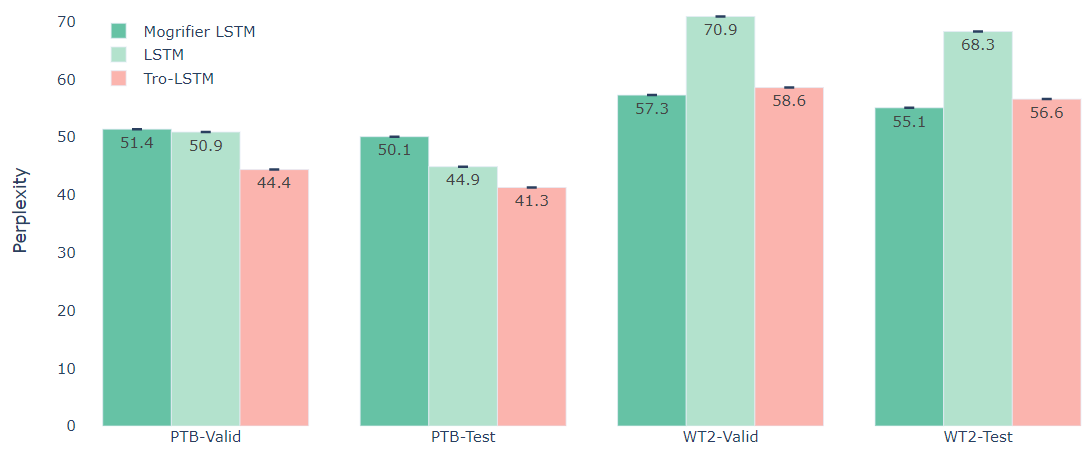}
  \caption{Perplexity of Language Models evaluated on datasets PTB and WT2. The Troenpy weighting based model Tro-LSTM achieves substantial perplexity reduction compared with the baseline model LSTM. The Mogrifier LSTM, a recent state-of-the-art, is included for reference. }
\end{figure}

\begin{table}[p]
 \caption{Top 10 Troenpy Ranked Words for PTB and WT2. The top WT2 words have much higher self-troenpy values than PTB, suggesting they are more characteristic influencing their neighboring words.}
 \centering
\begin{tabular}{ |p{3cm}| p{3cm}|| p{3cm}|p{3cm}| }
    \hline
    \multicolumn{2}{|c||}{PTB} & \multicolumn{2}{|c|}{WT2} \\
    \hline
    Word & Troenpy & Word & Troenpy \\
    \hline
    acceptances & 0.47 & guiry & 1.73 \\
    eurodollars & 0.46 & pld & 1.61\\
    adjustable & 0.20 &  managerial & 0.69 \\
    multiples & 0.18 &   włodzimierz & 0.47\\
    libor & 0.18 & juliusz & 0.36\\
    prebon & 0.16 & ftáčnik & 0.34\\
    exercisable & 0.13 & ginczanka & 0.34\\
    fulton & 0.12 & nd4 & 0.33\\
    capped & 0.10 & maharaj & 0.33\\
    ginnie & 0.10 & zuzanna & 0.31\\
    \hline
\end{tabular}

\end{table}

\section*{Quantum Troenpy for Quantum Information Theory of Certainty}
For a quantum system, the uncertainty includes  not only the classical uncertainty  but also the quantum uncertainty due to the entanglement in a quantum state. The associated density operator captures both types of uncertainty and gives a way to measure the probability of the outcomes of any physical measurement for the system. Let $\rho$ denote the density matrix, the Von Neumann entropy \cite{vonNeumann} is given as $H(\rho)=-\textrm{tr}(\rho\textrm{ln}\rho)$, where tr denotes the trace. Using the eigen-decomposition  of $\rho=\sum_j\eta_j|j><j|$,  where $|j>$ is a basis eigenvector corresponding to the eigenvalue $\eta_j$,  then it can also given as $H(\rho)=-\sum_j\eta_j\textrm{ln}{\eta_j}$. Note the $\eta_j$'s are probabilities with  sum 1, so it can be viewed as the classical Shannon entropy. Nowadays the quantum information theory \cite{wilde2017} is largely concerned with the uses of Von Neumann entropy, much as the classical information theory is mainly concerned with the interpretation and use of Shannon entropy.

Analogue to our discussion above, we can define a  corresponding {\bf quantum troenpy} to quantify the certainty of the quantum system as the usable quantum information for mathematical modeling and prediction purposes. Specifically we define
 $\textrm{T}(\rho):=\textrm{tr}(\rho\textrm{ln}(\textrm{I}-\rho))$, where $\textrm{I}$ is the identity matrix. It also equals the troenpy defined above $\textrm{T}(\rho)=-\sum_j\eta_j\textrm{ln}(1-\eta_j)$ in the sense of classical information theory. The corresponding properties can be established as analogue to the quantum Shannon information theory and the above proposal of troenpy for classical and non-quantum certainty. This proposal is obviously out of the scope of this paper and we will investigate on that  elsewhere.

\section*{Discussion}

Recently a few methods for self-supervised representation learning train feature extractors by maximizing an estimate of mutual information between different views of the training data, and have achieved some success 
in practice \cite{poole19a, TianCPC2019, Oord2018cpc, hjelm2018MI,  BachmanMIviews2019}. Unfortunately it has been pointed out that these success cannot be attributed to mutual information alone and they strongly depend on the feature extractor architectures and the parametrization of the mutual information estimator \cite{Michael2019MI}. This also suggests the current Shannon information theory may not be ideal for the unsupervised  representation learning and some new notion of information which can be usable is desired. Several researcher groups have proposed some computational methods to achieve such goals. For example, Xu et al \cite{Xu2020Usable} introduced predictive $\nu$-information which takes into account the modeling power, and Kleinman et al \cite{Kleinman2020UsableInfo} introduced usable information contained in the representation learned by deep networks to study the optimal representation during training. Our approach is more classical in nature  and easy to understand and implement. It integrates perfectly with Shannon information theory and it offers a dual view of the underlying data. A more profound impact is that since the troenpy is the desired usable information for machine learning, researchers may design new machine learning procedures and algorithms where the information extraction and maximization can be incorporated in the optimization process.

The discovery of troenpy reveals again that symmetry seems universal in our nature, even when we quantify the information. The crucial philosophy guiding this quick, smooth and pleasant discovery journey is that if a theory or method works, then there is a good chance that the reciprocal of the theory or method may also works, though not guaranteed. If the reciprocal does not work out as expected, that is also a good news and it suggests that the original theory or method fails to reveal some unknown mechanism for the problem. The philosophy we believed, explored and observed during the research investigation is very well stated in an ancient Chinese philosophy book about change, the I Ching\cite{1967ching}.

\section*{Datasets and Supplements}
Here we give the basic descriptions of the datasets used in the carried experiments.

{\bf Text Documents Datasets for Classification }
Here we use the popular seven open source benchmark datasets which has recently been used for evaluating the document classification algorithms such as optimal transportation based word moving distance method\cite{kusner2015doc}. For the data preprocessing and experiment settings we closely follow the setup configurations of Mikhail et al \cite{Mikhail2019}.
Specifically we use the datasets  to evaluate the models under the KNN based classification framework. The datasets include BBC sports news articles labeled into five sports categories (BBCsports);  medical documents labeled into 10 classes of cardiovascular disease types( Ohsumed); Amazon reviews labeled by three categories of Positive, Neutral and Negative (Amazon);  tweets labeled by sentiment categories (Twitter);  newsgroup articles labeled into 20 categories (20 News group);  sentences from science articles labeled by different publishers ( Classic) and Reuters news articles labeled by eight different topics (R8). The more detailed information about the datasets can be  found in the references mentioned above. For the datasets with no explicit train and test splits, we use the common 80/20 train-test split and report the performance result based on 50 repeats of random sampling. For easy comparison, we fix the neighborhood K=7 in all the KNN experiment settings.

{\bf Datasets for Language Modeling}
Two well-known benchmark datasets are used for the evaluation of language models at the word level. The Penn TreeBank (PTB) corpus by Marcus et al\cite{marcus1993}  has been long used as a benchmark dataset for language modeling and we use the preprocessed version by  Mikolov et al \cite{Mikolov2010}. The other dataset is the  Wikitext-2 (WT2) by Merity et al\cite{wt2}. The PTB dataset has about 10,000 unique words and it does not contain capital letters, numbers or punctuation.  The WT2 is sourced from Wikipedia and is approximately twice the size of PTB dataset. It has a vocabulary about 30,000 words. The text contains capitalization, punctuation and numbers. So WT2 is slightly noisy and hard to model.

\section*{Acknowledgments}
AJZ thanks Yingqing Huang, Tao Wu and Song Guo for their encouragement during the project. The related USPTO Patents 18161067 and 18304724 are pending.



\bibliography{scibib}

\bibliographystyle{Science}



\end{document}